%% file: main.tex
\documentclass[10pt, conference, compsocconf]{IEEEtran}

\usepackage{graphicx}
\usepackage{subfig}
\usepackage{tikz}
\usetikzlibrary{shapes.misc, positioning,calc}
\usetikzlibrary{fit}
\usepackage{amsmath}
\usepackage{multirow}
\usepackage{booktabs}
\usepackage{ifthen}
\usepackage{color}
\usepackage{url}
\usepackage{algorithmic}
\algsetup{indent=1em}
\usepackage{varwidth}
\usepackage{moresize}

\ifdefined\textln\relax\else\newcommand{\textln}[1]{#1}\fi
\ifdefined\shortcite\relax\else\newcommand{\shortcite}[1]{\cite{#1}}\fi
\ifdefined\nd\relax\else\newcommand{\nd}[1]{#1D}\fi

\newcommand{\pairs}[2]{\langle #1, #2 \rangle}

\definecolor{rd} {rgb} {1.0,0.0,0.0}
\definecolor{lg} {rgb} {0.7,0.3,0.9}
\definecolor{db} {rgb} {0.0,0.0,0.7}
\definecolor{dg} {rgb} {0.0,0.7,0.0}
\definecolor{dr} {rgb} {0.7,0.0,0.0}
\definecolor{gr} {rgb} {0.8,0.4,0.1}

\newcommand{\TODO  } [1] {{{\color{rd}[TODO] #1}}}
\newcommand{\John  } [1] {{{\color{db}[John] #1}}}
\newcommand{\Saman } [1] {{{\color{dg}[Saman]  #1}}}
\newcommand{\Martin} [1] {{{\color{lg}[Martin]#1}}}
\newcommand{\revision} [1] {#1}
\newcommand{\remove} [1] {#1}
\newcommand{\replace}[2]{#2}

\newcommand{\final}{0}

\ifthenelse{\equal{\final}{1}} {\renewcommand{\TODO  }[1]{}}{}
\ifthenelse{\equal{\final}{1}} {\renewcommand{\John  }[1]{}}{}
\ifthenelse{\equal{\final}{1}} {\renewcommand{\Saman }[1]{}}{}
\ifthenelse{\equal{\final}{1}} {\renewcommand{\Martin}[1]{}}{}

\begin{document}

\title{A Dynamic Hash Table for the GPU}


\author{%
\IEEEauthorblockN{Saman Ashkiani}
\IEEEauthorblockA{Electrical and Computer Engineering\\
University of California, Davis\\
sashkiani@ucdavis.edu}
\and
\IEEEauthorblockN{Martin Farach-Colton}
\IEEEauthorblockA{Computer Science\\
Rutgers University\\
farach@cs.rutgers.edu}
\and
\IEEEauthorblockN{John D. Owens}
\IEEEauthorblockA{Electrical and Computer Engineering\\
University of California, Davis\\
jowens@ece.ucdavis.edu}
}

\maketitle

\input{abstract}
\graphicspath{{}}
\input{introduction.tex}
\input{related.tex}
\input{linked_list.tex}
\input{impl_details.tex}
\input{mem_alloc.tex}
\input{perf_eval.tex}
\input{conclusion.tex}
\input{ack.tex}
\let\doi\relax                  % otherwise DOIs won't print
\bibliographystyle{IEEEtran}
\bibliography{all,temp}

\end{document}

%% file: abstract.tex
\begin{abstract}
We design and implement a fully concurrent dynamic hash table for GPUs with comparable performance to the state of the art static hash tables.
We propose a warp-cooperative work sharing strategy that reduces branch divergence and provides an efficient alternative to the traditional way of per-thread (or per-warp) work assignment and processing.
By using this strategy, we build a dynamic non-blocking concurrent linked list, the \emph{slab list}, that supports asynchronous, concurrent updates (insertions and deletions) as well as search queries.
We use the slab list to implement a dynamic hash table with chaining (the \emph{slab hash}).
On an NVIDIA Tesla K40c GPU, the slab hash performs updates with up to 512~M updates/s and processes search queries with up to 937~M queries/s.
We also design a warp-synchronous dynamic memory allocator, \emph{SlabAlloc}, that suits the high performance needs of the slab hash.
SlabAlloc dynamically allocates memory at a rate of 600~M allocations/s, which is up to 37x faster than alternative methods in similar scenarios.
\end{abstract}

%% file: introduction.tex
\section{Introduction}\label{sec:intro}
A key deficiency of the GPU ecosystem is its lack of \emph{dynamic} data structures, which allow incremental updates (such as insertions and deletions). Instead, GPU data structures (e.g., cuckoo hash tables~\cite{Alcantara:2009:RPH:nourl}) typically address incremental changes to a data structure by rebuilding the entire data structure from scratch. A few  GPU data structures (e.g., the dynamic graph data structure in cuSTINGER~\cite{Green:2016:CSD}) implement phased updates, where updates occur in a different execution phase than lookups.  In this work we describe the design and implementation of a hash table for GPUs that supports truly concurrent insertions and deletions that can execute together with lookups.

Supporting high-performance concurrent updates of data structures on GPUs represents a significant design challenge. Modern GPUs support tens of thousands of simultaneous resident threads, so traditional lock-based methods that enforce concurrency will suffer from substantial contention and will thus likely be inefficient.
Non-blocking approaches offer more potential for such massively parallel frameworks, but most of the multi-core system literature (e.g., classic non-blocking linked lists~\cite{Michael:2002:HPD}) neglects the sensitivity of GPUs to memory access patterns and branch divergence,
which makes it inefficient to directly translate those ideas to the GPU\@.

In this paper, we present a new GPU hash table, the \emph{slab hash}, that supports bulk and incremental builds.  One might expect that supporting incremental insertions and deletions would result in significantly reduced query performance compared to static data structures.  However, our hash table not only supports updates with high performance but also sustains build and query performance on par with static GPU hash tables. Our hash table is based on a novel linked list data structure, the \emph{slab list}. Previous GPU implementations of linked lists~\cite{Misra:2012:PEC}, which operate on a thread granularity and contain a data element and pointer per linked list node, exhibit poor performance because they suffer from control and memory divergence and incur significant space overhead. The slab list instead operates on a warp granularity, with a width equal to the SIMD width of the underlying machine and contains many data elements per linked list node. Its design minimizes control and memory divergence and uses space efficiently. We then construct the slab hash from this slab list as its building block, with one slab list per hash bucket. Our contributions in this work are as follows:

\begin{itemize}
  \item The slab list is based on a node structure that closely matches the GPU's hardware characteristics.
  \item The slab list implementation leverages a novel warp-cooperative work sharing strategy that minimizes branch divergence, using warp-synchronous programming and warp-wide communications.
  \item The slab hash, based on the slab list, supports concurrent operations with high performance.
  \item To allow concurrent updates, we design and implement a novel memory allocator that dynamically and efficiently allocates and deallocates memory in a way that is well-matched to our underlying warp-cooperative implementation.
  \item Our memory allocator is scalable, allowing us to support data structures up to 1~TB (far larger than the memory size of current GPUs) and without any CPU intervention.
  \item The slab hash's bulk-build and search rates are comparable to those of static methods (e.g., GPU cuckoo hashing~\cite{Alcantara:2009:RPH:nourl}), while additionally achieving efficient incremental updates.
\end{itemize}

%% file: related.tex
\section{Background \& Related Work}\label{sec:related}
\remove{
\paragraph{Graphics Processing Unit (GPU)}
GPUs are massively parallel processors with thousands of parallel active threads.
Threads are grouped into SIMD units of width 32---a \emph{warp}---and each warp executes instructions in lockstep.
As a result, any branch statements that cause threads to run different instructions are serialized (branch divergence).
A group of threads (multiple warps) are called a thread block and are scheduled to be run on different streaming processors (SMs) on the GPU.
The memory hierarchy of GPUs is organized into a large global memory accessible by all threads within the device (e.g., 12~GB on the Tesla K40c), smaller but faster shared memory for each thread block (48~KB per SM on the Tesla K40c), and local registers for each thread in the thread block (64~KB per SM on the Tesla K40c).
Maximizing achieved memory bandwidth requires accessing consecutive memory indices within a warp (\emph{coalesced} access).
NVIDIA GPUs support a set of warp-wide instructions (e.g., shuffles and ballots) so that all threads within a warp can communicate with each other.
}
% %%%% linked lists and skip lists:
% \paragraph{Linked lists}
% \John{What kind of mutations do these support?}
% Misra and Chaudhuri implemented a classic non-blocking linked list as well as a skip list for GPUs. Their implementation does not do anything special to avoid GPU memory and control divergence~\cite{Misra:2012:PEC}.
% Recently, Moscovici et al.\ proposed a lock-based GPU-friendly skip list (GPSL) with an emphasis on GPU's preferred coalesced memory accesses~\cite{Moscovici:2017:PGS}.
% In the best case and without considering any pointer dereferencing, an update would require two atomic operations (for handling the lock) and two regular coalesced memory accesses (read and write).
% Searching for an element would require two regular memory accesses. \Martin{I don't understand what I'm supposed to conclude for this last part.  Is that good or bad?}
% Would reflect these in Performance Evaluation Section:
% Peak performance on a GeForce GTX 970, with 224 GB/s memory bandwidth, is reported to be up to 100~M queries/s for searches and 50~M updates/s for insertion and deletions.

%%%% hash tables:
\paragraph{Hash tables}
There are several efficient static hash tables implemented for GPUs.
Alcantara et al.~\cite{Alcantara:2009:RPH:nourl} proposed an open-addressing cuckoo hashing scheme for GPUs. This method supports bulk build and search,  both of which require minimal memory accesses in the best case: a single atomic operation for inserting a new element, and a regular memory read for a search.
As the load factor increases, it is increasingly likely that a bulk build using cuckoo hashing fails.
Garcia et al.~\cite{Garcia:2011:CPH} proposed a method based on Robin Hood hashing that focuses on higher load factors and uses more spatial locality for graphics applications, at the expense of performance degradation compared to cuckoo hashing.
Khorasani et al.'s stadium hashing is also based on a cuckoo hashing scheme but stores two tables instead of one.
Its focus is mainly on out-of-core hash tables that cannot be fit on a single GPU's memory.
In the best case (i.e., an empty table) and with a randomly generated key, an insertion in this method requires one atomic operation and a regular memory write.
A search operation in stadium hashing requires at least two memory reads.
Although hash tables may be specifically designed for special applications, Alcantara's cuckoo hashing appears to be the best general-purpose in-core hash table option with the best performance measures. We use this method for our comparisons in Section~\ref{sec:perf_eval}.

Misra and Chaudhuri~\cite{Misra:2012:PEC} implemented a lock-free linked list, which led to a lock-free hash table with chaining that supported concurrent insertion, deletion and search.
Their implementation is not fully dynamic, because it pre-allocates all future insertions into an array (which must be known at compile time), and it does not address the challenge of dynamically allocating new elements and deallocating deleted elements at runtime. However, we briefly compare it to the slab hash in Section~\ref{subsec:concurrent}.
\remove{Inspired by Misra and Chaudhuri's work, Moscovici et al.~\cite{Moscovici:2017:PGS} recently proposed a lock-based GPU-friendly skip list (GFSL) with an emphasis on the GPU's preferred coalesced memory accesses.
We will also discuss in Section~\ref{subsec:concurrent} why we believe GFSL (either by itself or as a building block of a larger data structure) cannot outperform our lock-free slab hash in updates and searches.}  I/O sensitive linked lists were studied in the CPU context by Bender et al.~\cite{Bender:2002:STM}.

%%%% dynamic memory allocation
\paragraph{Dynamic memory allocation} Although a mature technology for single and multi-core systems, dynamic memory allocation is still considered a challenging research problem on massively parallel frameworks such as GPUs.
Massive parallelism makes it difficult to directly exploit traditional allocation strategies such as lock-based or private-heap schemes without a significant performance degradation.

CUDA~\cite{CUDA:2016}  provides a built-in \texttt{malloc} that dynamically allocates memory on the device (GPU).
However, it is not efficient for small allocations (less than 1~kB).
To address \texttt{malloc}'s inefficiencies for small allocations, almost every competitive proposed method so far is based on the idea of allocating numerous large enough memory pools (with different terminology), assigning each memory pool to a thread, a warp, or a thread block (to decrease parallel contention), dynamically allocating or deallocating small portions of it based on received requests, and finally implementing a mechanism to use another memory pool once fully allocated.
Some methods use hashing to operate on different memory pools (e.g.,
% \texttt{ScatterAlloc}~\cite{Steinberger:2012:SMP} and
\texttt{Halloc}~\cite{Adinetz:2014:HHT}).
Other methods use various forms of linked lists to move into different memory pools (e.g.,
% \texttt{XMalloc}~\cite{Huang:2010:XSL} and
\texttt{CMalloc}~\cite{Vinkler:2015:RED}).
All these methods maintain various flags (or bitmaps) and operate on them atomically to be able to allocate or deallocate memory.

Vinkler et al.\ has provided an extensive study of all these methods and some benchmarks to compare their performance~\cite{Vinkler:2015:RED}.
The most efficient ones, \texttt{CMalloc} and \texttt{Halloc}, perform best when there are multiple allocation requests within each warp that can be formed into a single but larger allocation per warp (a \emph{coalesced} allocation).
\revision{However, for the warp-cooperative work sharing strategy we use in this work (Section~\ref{subsec:warp_focus}), we need an allocator that can handle numerous independent but sequentially available allocation requests per warp, which cannot be formed into a single larger coalesced allocation to avoid divergence overheads.} As we will see in Section~\ref{sec:mem_alloc}, existing allocators perform poorly in such scenarios.
Instead, we propose a novel warp-synchronous allocator, \emph{SlabAlloc}, that uses the entire warp to efficiently allocate fixed-size slabs with modest register usage and minimal branch divergence (more details in Section~\ref{sec:mem_alloc}).

% \John{A couple of sentences on warp-synchronous programming? You can also provide a forward pointer to Section~\ref{subsec:warp_focus}.}

% \John{Put your work in context with this previous work! I think that your primary goal is performance and you should say that. And what is different about your design than the ``almost every competitive proposed method''?}
% \texttt{FDGMalloc} exclusively maintains these memory pools by different warps to avoid further contentions~\cite{Widmer:2013:FDM}.

% \begin{itemize}
%       \item background on linked lists on CPU and GPU
%       \item background on different hash table works
% \end{itemize}

%% file: linked_list.tex
\section{Design description}\label{sec:design}
A linked list is a linear data structure whose elements are stored in non-contiguous parts of the memory.
These arbitrary memory accesses are handled by storing the memory address (i.e., a pointer) of the next element of the list alongside the data stored at each node.
The simplicity of linked lists makes concurrent updates relatively easy to support, using compare-and-swap (CAS) operations~\cite{Michael:2002:HPD}.
New nodes can be inserted by (1) allocating a new node, (2) initializing the data it contains and storing the successor's address into its \emph{next} pointer, then (3) atomically compare-and-swapping the new node's address with its predecessor's next pointer.
Similarly, nodes can be deleted by (1) atomically marking a node as deleted (to make sure no new node is inserted beyond it) and then (2) compare-and-swapping its predecessor's pointer with its successor's address.

On GPUs, it is possible to implement the same set of operations for a linked list, and then use it as a building block of other data structures (e.g., in hash tables)~\cite{Misra:2012:PEC}.
However, this implementation requires an arbitrary random memory access per unit of stored data, which is not ideal for any high-performance GPU program.
Furthermore, making any change to a linked list data structure requires dynamic memory allocation, which itself is challenging to perform efficiently, especially on massively parallel devices such as GPUs (Section~\ref{sec:related}).
In this work, we propose a new linked list data structure, the slab list, and then use it to implement a dynamic hash table (slab hash). In our design, we have two major goals in mind: (1) maximizing performance in maintaining several slab lists concurrently (suited for hash tables), and (2) having better memory utilization by reducing the memory overhead in classic linked lists.
We present the slab list and slab hash in this section, and then provide implementation details in Section~\ref{sec:impl_details}.
%%%%%%%%%%%%%%%%%%%%%%%%%%%%%%%%%%%
\subsection{Slab list}\label{subsec:linked_list}
Classic singly linked lists consist of nodes with two main distinctive parts: a single unit of data element (a key, a key-value pair, or any other arbitrary metadata associated with a key), and a \emph{next} pointer to its successor node.
Storing the successor's memory address makes linked lists flexible and powerful in dealing with mutability issues. However, it introduces additional memory overhead per stored unit of data.
Moreover, the efficiency of linked list operations is one of our primary concerns.

In classic linked list usage, an operation (insertion/deletion/search) is often requested from a single independent thread. These high-level operations translate into lower-level operations on the linked list itself. In turn, these lower-level operations result in a series of random, sequential memory operations per thread.
Because we expect that a parallel program that accesses such a data structure will feature numerous simultaneous operations on the data structure, we can easily parallelize operations across threads.
But in modern parallel hardware with SIMD cores, including GPUs, the peak memory bandwidth is only achieved when threads within each SIMD unit (i.e., a warp in GPUs) access consecutive memory indices with a certain fixed alignment (e.g., on NVIDIA GPUs, each thread fetches a 32-bit word per memory access, i.e., 128~bytes per warp).

There are some well-known tactics to avoid coalesced-memory issues in GPUs, such as using structure-of-arrays instead of array-of-structures data layouts, or first fetching a big block of items into a faster but locally shared memory (with coalesced memory accesses) and then accessing the local memory with an arbitrary alignment.
However, none of these methods is effective with a linked list data structure that requires singleton structures distributed randomly in the memory domain.
As a result, we propose to use an alternate linked list design that is more suitable for our hardware platform.

\begin{figure}
\input{memory_units.tex}
\caption{Regular linked list and the slab list.}
\vspace{-0.2in}
\label{fig:memory_units}
\end{figure}
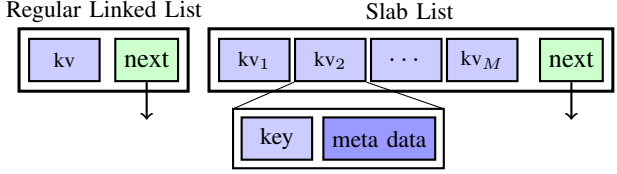

\revision{
Braginsky and Petrank proposed a lock-free linked list on the CPU~\cite{Braginsky:2011:LCL} that achieves better locality of reference by ensuring that certain number of regular linked list nodes (an \emph{entry}; a data and a pointer) will all be arbitrarily placed into a larger contiguous structure (a \emph{chunk}) that would fit in a single cache line. As a result, each entry would only point to another entry within that chunk. Each chunk itself would point to another chunk to form the whole list.
We also achieve better locality, but in a different way. We use a larger linked list node (called a \emph{slab}, or interchangeably a \emph{memory unit}) that consists of multiple data elements and a single pointer to its successor slab (shown in Fig.~\ref{fig:memory_units}).
The main difference is that slabs are fixed in size, and all data elements within a slab share a single next pointer.
}

An immediate advantage of slab lists is that their memory overhead is reduced by approximately a factor of $M$ (if there are $M$ data elements per slab).
However, our main motivation for using large slabs is to be able to maintain them in parallel, meaning that the whole slab is accessed with a minimum number of memory accesses and in parallel (distributed among multiple threads), and then operations are also performed in parallel.
The optimal size of these slabs will depend on the hardware characteristics of the target platform, including both the way memory accesses are handled as well as communication possibilities among different threads.

On GPUs, we operate on each slab with a single SIMD unit (a warp) and use  available warp-wide intrinsics such as shuffles and ballots for communications. So, the size of a slab will be a modest multiple of the warp width (e.g., 32 consecutive 32-bit words).
\revision{We do not maintain order within our slabs. GPU hardware enables us to search within an unordered set of 32 words with a single ballot instruction. So, as long as we keep the slab list relatively short (e.g., $\sim$10 slabs), we can have faster updates with negligible extra search cost. If not, extra measures should be taken, including maintaining an inter-slab order.}

%%%%%%%%%%%%%%%%%%%%%%%%%%%
\subsection{Supported Operations in Slab Lists}\label{subsec:operations}
Suppose our slab list maintains a set of keys (or key-value pairs), here represented by $\mathcal{S}$. Depending on whether or not we allow duplicate keys in our data structure, we support the following operations:
\begin{itemize}
        \item \textsc{insert}$(k,v)$: $\mathcal S \leftarrow \mathcal S \cup \{ \pairs{k}{v}\}$. Insert a new key-value pair into the slab list.
        \item \textsc{replace}$(k,v)$: $\mathcal{S} \leftarrow (\mathcal S  -  \{ \pairs{k}{*} \}) \cup \{ \pairs{k}{v}\}$. Insert a new key-value pair with an extra restriction on maintaining uniqueness among keys (i.e., replace a previously inserted key if it exists).
        \item \textsc{delete}$(k)$: $\mathcal{S} \leftarrow \mathcal{S} - \{\pairs{k}{v} \in \mathcal{S}\}$. Remove the least recently inserted key-value pair $\pairs{k}{v}$.
        \item \textsc{deleteAll}$(k)$: $\mathcal{S} \leftarrow \mathcal S - \{ \pairs{k}{*}\}$. Delete all instances of a key in the slab list.
        \item \textsc{search}$(k)$: Return the least recent $\pairs{k}{v} \in \mathcal{S}$, or $\perp$ if not found.
        \item \textsc{searchAll}$(k):$ Return all found instances of $k$ in the data structure ($\{\pairs{k}{*} \in \mathcal{S}\}$), or $\perp$ if not found.
\end{itemize}
\subsubsection{Search (\textsc{search} and \textsc{searchAll})} Searching for a specific key in slab list is similar to classic linked lists. We start from the head of the list and look for the key within that memory unit. If none of the data units possess such a key, we load the next memory unit based on the stored successor pointer.
In \textsc{search} we return the first found matching element, but in \textsc{searchAll} we continue searching the whole list. In both cases, if no matching key is found we return $\perp$.

\subsubsection{Insertion (\textsc{insert} and \textsc{replace})}
For the \textsc{insert} operation, we make no extra effort to ensure uniqueness among the keys, which makes the operation a bit easier. We simply start from the head of the list and use an atomic CAS to insert the new key-value pair into the first empty data unit we find.
If the CAS operation is successful, then insertion is done.
Otherwise, it means that some other thread has successfully inserted a pair into that empty data unit, and we have to try again and look for a new empty spot.
If the current memory unit is completely full (no empty spot), we load the next memory unit and repeat the procedure.
If we reach the end of the list, it means that the linked list requires more memory units to contain the new element.
As a result, we dynamically allocate a new memory unit and use another atomic CAS to replace the null pointer currently existing in the tail's successor address with the address of the newly allocated memory unit.
If it is successful, we restart our insertion procedure from the tail again.
If it failed, it means some other thread has successfully added a new memory unit to the list.
Hence, we release the newly allocated memory unit and restart our insertion process from the tail.

\textsc{Replace} is similar to \textsc{insert} except that we have to search the entire list to see if there exists a previously inserted key $k$. If so, then we use atomic CAS to replace it with the new pair.
If not, we simply perform \textsc{insert} starting from the tail of the list.

\subsubsection{Deletion (\textsc{delete} and \textsc{deleteAll})}
\revision{
To delete a key, we start from the head slab and look for the matching key. If found, we mark the element as deleted.\footnote{In our design we reserve two 32-bit values in the key domain to denote 1) an empty spot, and 2) a deleted key.}
If not, we continue to the next slab.
We continue this process until we reach the end of the list.
For \textsc{delete}, we return after deleting the first matching element, but for \textsc{deleteAll} we process the whole list.
We later describe our \textsc{flush} operation that locks the list and removes all stale elements (marked as deleted) and rebalances the list to have the minimum number of necessary memory units, releasing extra memory units for later allocations.

In case we allow duplicates, we can simply mark a to-be-deleted element as empty. In this case, later insertions that use \textsc{insert} can potentially find these empty spots down the list and insert new items in them.
However, if we do not allow duplicates, in order to correctly maintain the uniqueness condition, we must mark deleted elements differently than being empty to avoid inserting a key that already exists in the list (somewhere in its successive memory units).
}
% \John{Seems to me it would be good to have a couple of paragraphs that describe the ability of this data structure to do truly concurrent operations. \texttt{paragraph} would be good.}\Saman{Will come back to this soon.}
%%%%%%%%%%%%%%%%%%%%%%%%%%%%%%%%%%%
\input{hash_table.tex}
% \subsection{Dynamic Hash Table}\label{subsec:hash_table}

%% file: memory_units.tex
\begin{tikzpicture}[%
auto,
  data_block/.style={
    rectangle,
    draw=black,
    thick,
    fill=blue!20,
    text width=2.0em,
    align=center,
    minimum height=1.6em
  },
  next_block/.style={
    rectangle,
    draw=black,
    thick,
    fill=green!20,
    minimum width = 2.0em,
    align=center,
    minimum height=1.6em
  },
  meta_block/.style={
    rectangle,
    draw=black,
    thick,
    fill=blue!40,
    minimum width = 2.0em,
    align=center,
    minimum height=1.6em
  }    
]
\def \offsetDataX {0.25}
\def \offsetDataY {0.25}
\def \dx {0.25}
\def \DX {1}

% drawing regular linked lists:
\draw [black] (\offsetDataX, \offsetDataY) node[data_block] (A) {\footnotesize kv};
\draw [black] (\offsetDataX + 3.8em, \offsetDataY) node[next_block] (B) {next};
\draw [black, line width = 0.4mm] ($(A.north west) + (-0.12, 0.12)$) rectangle ($(B.south east) + (0.12, -0.12)$);

\draw [black, thick, ->] ($(B.south)$) -- ($(B.south) + (0,-0.5)$);
\draw [] ($(A.north) + (1.5em, 1em)$) node{\small Regular Linked List};
%%%% drawing our memory units:

\draw [black] ($(B.east) + (1.0, 0)$) node[data_block] (A1) {\footnotesize $\text{kv}_1$};
\draw [black] ($(A1.east) + (1.5em, 0)$) node[data_block] (A2) {\footnotesize $\text{kv}_2$};
\draw [black] ($(A2.east) + (1.5em, 0)$) node[data_block] (A3) {$\dots$};
\draw [black] ($(A3.east) + (1.5em, 0)$) node[data_block] (A4) {\footnotesize $\text{kv}_{M}$};

\draw [black] ($(A4.east) + (2em, 0)$) node[next_block] (B1) {next};
\draw [black, line width = 0.4mm] ($(A1.north west) + (-0.12, 0.12)$) rectangle ($(B1.south east) + (0.12, -0.12)$);
\draw [black, thick, ->] ($(B1.south)$) -- ($(B1.south) + (0,-0.5)$);
\draw [] ($(A2.north) + (3em, 1em)$) node{\small Slab List};

%%% magnified:
\draw [black] ($(A1.south) + (0.3, -0.75)$) node[data_block] (C1) {\small key};
\draw [black] ($(C1.east) + (2.5em, 0)$) node[meta_block] (C2) {\small meta data};
\draw [black, thick] ($(C1.north west) + (-0.10, 0.10)$) rectangle ($(C2.south east) + (0.10, -0.10)$);
\draw [black] ($(A2.south west)$) -- ($(C1.north west) + (-0.10, 0.10)$);
\draw [black] ($(A2.south east)$) -- ($(C2.north east) + (0.10, 0.10)$);
\end{tikzpicture}

%% file: hash_table.tex
\subsection{Slab Hash: A Dynamic Hash Table}\label{subsec:hash_table}
\revision{Our \emph{slab hash} is a \emph{dynamic} hash table (meaning that we support not only operations like searches that do not change the contents of the hash table but also operations like insertions and deletions that do) built from a set of $B$ slab
lists (buckets).\footnote{\revision{Similar to other hash tables (on CPU or GPU), the slab hash is capacity based, meaning that our performance depends on the initial number of buckets. As shown in Section~\ref{sec:perf_eval}, for any choice of $B$, we can cause performance degradation by continually increasing the number of elements (but it never breaks).}} This hash table uses chaining as its collision resolution.}
More specifically, we use a direct-address table of $B$ buckets (\emph{base slabs}), where each bucket corresponds to a unique hashed value from $0$ to $B-1$~\cite{CLRS:2005:ITA}.
Each base slab is the head of an independent slab list consisting of slabs, as introduced in Section~\ref{subsec:linked_list}, each with $M$ data points to be filled.
In general, base slabs and regular slabs can differ in their structures in order to allow additional implementation features (e.g., pointers to the tail, number of slabs, number of stored elements, etc.).
For simplicity and without loss of generality, here we assume there is no difference between them.

We use a simple universal hash function such as $h(k; a, b) = ((ak + b) \bmod p) \bmod B$, where $a, b$ are random arbitrary integers and $p$ is a random prime number.
As a result, on average, keys are distributed uniformly among all buckets with an \emph{average slab count} of $\beta = n/(MB)$ slabs per bucket, where $n$ is the total number of elements in the hash table.
For searching a key that does not exist in the table (i.e., an unsuccessful search), we should perform $\Theta(1 + \beta)$ memory accesses.
A successful search is slightly better, but has similar asymptotic behavior.

In order to be able to compare our memory usage with open-addressing hash tables that do not use any pointers (e.g., cuckoo hashing~\cite{Alcantara:2009:RPH:nourl}), we define the \emph{memory utilization} to be the amount of memory actually used to store the data over the total amount of used memory (including pointers and unused empty slots).
If each element and pointer take $x$ and $y$ bytes of memory respectively, then each slab requires $Mx + y$ bytes.
\revision{As a result, our slab hash would achieve a memory utilization equal to $\frac{x}{Mx+y}\frac{n}{\sum_{i=0}^{B-1}{k_i}} \leq \frac{Mx}{Mx+y}$, where $k_i$ denotes the number of slabs for bucket $i$.}
% In order to achieve a certain memory utilization given a particular slab structure, we compute the required $\beta$, which is directly affected by the total number of buckets in the hash table ($\beta = n/(MB)$).
For open-addressing hash tables, memory utilization is equal to the load factor, i.e., the number of stored elements divided by the table size.

%% file: impl_details.tex
\section{Implementation details}\label{sec:impl_details}
\remove{In this section we focus on our technical design choices and implementation details, primarily influenced by the hardware characteristics of NVIDIA GPUs.}
%%%%%%%%%%%%%%%%%%%%%%%%%%%%%%%
\subsection{Our warp-cooperative work sharing strategy}\label{subsec:warp_focus}
A traditional, but not necessarily efficient, way to perform a set of independent tasks on a GPU is to assign and process an independent task on each thread (e.g., classic linked list operations on GPU~\cite{Misra:2012:PEC}).
An alternative approach is to do a per-warp work assignment followed by a per-warp processing (e.g., warp-wide histogram computation~\cite{Ashkiani:2016:GM:nourl}).
In this work we propose a new approach where threads are still assigned to do independent tasks (per-thread assignment), but works are done in parallel (per-warp processing).
We call this a \emph{warp-cooperative work sharing (WCWS)} strategy.
\revision{This strategy would be particularly useful under the following circumstances: 1)~threads are assigned to independent-but-different tasks (irregular workload); 2)~each task requires an arbitrarily placed but vectorized memory access (accessing consecutive memory units); 3)~it is possible to process each task in parallel within a warp using warp-wide communication (warp friendly).}
In our data structure context, this means that we form a work queue of arbitrary requested operations from different threads within a warp, and all threads within that warp cooperate to process these operations one at a time (based on a pre-defined intra-warp order) and until the whole work queue is empty.

If data is properly distributed among the threads, as it is naturally in our slab based design, then regular data structure operations such as looking for a specific element can simply be implemented in parallel using simple warp-wide instructions (e.g., using ballots and shuffles).
An immediate advantage of the WCWS strategy is that it significantly reduces branch divergence when compared to traditional per-thread processing.
A disadvantage is that we should always keep all threads within a warp active in order to correctly perform even a single task (avoiding branches on threads). But, this limitation already exists on many CUDA warp-wide instructions and can be easily avoided by using the same tricks~\cite[Chapter B.15]{CUDA:2016}.
%%%%%%%%%%%%%%%%%%%%%%%%%%%%%%%
\subsection{Choice of parameters}\label{subsec:parameters}
As we emphasized in Section~\ref{sec:design}, the main motivation behind introducing slabs in our design is to have better coalesced memory accesses.
Hence, we chose our slab sizes to be a multiple of each warp's physical memory access width, i.e., at least 32$\times$4\,B for current architectures.
Throughout the rest of the paper, we assume each slab is exactly 128\,B, so that once a warp accesses a slab each thread has exactly 1/32 of the slab's content.
So, when we use the term ``lane'' for a slab, we mean that portion of the slab that is read by the corresponding warp's thread.
We currently support two item data types: 1) 32-bit entries (key-only), 2) 64-bit entries (key-value pairs), but our design can be extended to support other data types.
In both cases, slab lanes 0--29 contain the data elements (in the key-value case, even and odd lanes contain keys and values respectively). We refer to lane 31 as the \emph{address lane}, while lane 30 is used as an auxiliary element (flags and pointer information if required).
As a result, slab lists (and the derived slab hash) can achieve a maximum memory utilization of 94\%.

%%%%%%%%%%%%%%%%%%%%%%%%%%%%%%%
\subsection{Operation details}\label{subsec:operation_details}
Here we provide more details about some of slab hash operations discussed in Section~\ref{subsec:operations}.
\revision{We thoroughly discuss \textsc{search}, \textsc{replace} (insertion when uniqueness is maintained), and \textsc{delete}, and then briefly explain our methodology for the \textsc{flush} operation.}
In our explanations, we use some simplified code snippets. For example, ReadSlab() takes a 32-bit address layout of a slab as input; each thread reads its corresponding data portion.
SlabAddress() extends a 32-bit address layout to a 64-bit memory address (more details about memory address layouts are in Section~\ref{sec:mem_alloc}).

\subsubsection{\textsc{search}}\label{subsubsec:search_details}
Figure~\ref{alg:search} shows a simplified pseudocode for the \textsc{search} procedure in our slab hash.
As an input, any thread that has a search query to perform sets \texttt{is\_active} to true.
Keys are stored in \texttt{myKey} and the result will be stored in \texttt{myValue}.
By following the WCWS strategy introduced before, all threads within a warp participate in performing every search operation within that warp, one operation at a time.
% Describing the code in Figure alg:search
First, we form a local warp-wide work queue (line 3) by using a ballot instruction and asking whether any thread has something to search for.
Then, all threads go into a while loop (line 4) and repeat until all search queries are processed.
At each round, all threads can process the work queue and find the next lane within the warp that has the priority to perform its search query (the source lane, line 6).
This is done by using a pre-defined procedure \texttt{next\_prior()}, which can be implemented as simply as finding the first set bit in the work queue (using CUDA's \texttt{\_\_ffs}).
Then all threads ask for the source lane's query key using a shuffle instruction (line 6), and hash it to compute its corresponding bucket id (line 7).

The whole warp then performs a coalesced memory read from global memory (ReadSlab()), which takes the 32-bit address layout of the slab as well as the lane id of each thread.
If we are currently at the linked list's base slab (the bucket head), we will find the corresponding slab's contents in a fixed array.
Otherwise, we use our SlabAlloc allocator and compute the unique 64-bit address of that allocated slab by using the 32-bit \texttt{next} variable.
% Depending on whether we are currently at the linked list's base slab (bucket's head) or not, memory access may be different (line 9). For allocated slabs, we use our SlabAlloc allocator and the \texttt{address\_decode()} procedure that computes the unique 64-bit address of that allocated slab by using the 32-bit next variable (more details about our allocator are in Section~\ref{sec:mem_alloc}).
Now, every thread has read its portion of the target slab.
By using a ballot instruction we can ask whether any valid thread possesses the source lane's query (\texttt{src\_key}), and then compute its position \texttt{found\_lane} (line 14).
If found, we ask for its corresponding value by using a shuffle instruction and asking for its subsequent thread's \texttt{read\_data}, which stores the result from the requested source lane.
The source lane then stores back the result and marks its query as resolved (line 18).
If not found, we must go to the next slab. To find it, we ask the address lane for its address (line 21) and update the \texttt{next\_ptr}.
If the \texttt{next\_ptr} was empty, it means that we have reached the slab list's tail and the query does not exist (line 24). Otherwise, we update the \texttt{next} variable and continue within the next loop.
At each loop, we initially check whether the work queue has changed (someone has successfully processed its query) or we are still processing the same query (but are now searching in allocated slabs rather than the base slab).

%%%%
\input{code_example.tex}
%%%%

\subsubsection{\textsc{replace}}\label{subsubsec:insert_details}
The main skeleton of the \textsc{replace} procedure (Fig.~\ref{alg:search}) is similar to search, but now instead of looking for a particular key, we look for either that same key (to replace it), or an empty spot (to insert it for the first time).
Any thread with an insertion operation will mark \texttt{is\_active} as true.
As with search, threads loop until the work queue of all insertion operations are completely processed.
Within a loop, all threads read their corresponding portions of the target slab (lines 2--7), searching for the source key or an empty spot (an empty key-value pair) within the read slab (called the destination lane).
If found, the source lane inserts its key-value pair into the destination lane's portion of the slab with a 64-bit atomicCAS operation.
If that insert is successful, the source lane marks its operation as resolved, which will be reflected in the next work queue computation.
If the insert fails, it means some other warp has inserted into that empty spot and the whole process should be restarted.

If no empty spot or source key is found at all, all threads fetch the next slab's address from the address lane.
If that address is not empty, the new slab is read and the insertion process repeats.
If the address is empty, it means that a new slab should be allocated. All threads use the SlabAlloc routine, allocating a new slab, then the source lane  uses a 32-bit atomicCAS to update the empty address previously stored in the address lane.
If the atomicCAS is successful, the whole insertion process is repeated with the newly allocated slab.
If not, it means some other warp has successfully allocated and inserted the new slab and hence, this warp's allocated slab should be deallocated.
The process is then restarted again with the new valid slab.
%%%%%%%%%%%%%%%%%%%%%%%%%%%%%%%%%%%%%%%%
\subsubsection{\textsc{delete}}\label{subsubsec:delete_details}
\revision{
Deletion (shown in Fig.~\ref{alg:search}) is similar to both the \textsc{search} and \textsc{replace} operations. Each thread with a deletion operation to perform (a true \texttt{is\_active}) updates the work queue accordingly.
Then, for each deletion operation in the work queue, the source lane and its to-be-deleted key are queried by the whole warp (lines 2--7).
Now, the current slab is searched for the source key. We name the lane that possesses it as the destination lane (line 56).
If the destination lane is valid (a match is found), the source lane itself proceeds with overwriting the corresponding element with \texttt{DELETED\_KEY} (line 59).
If not found, then the next pointer is updated (line 63).
If we reach the end of the list (an empty next pointer), the source key does not exist in the list and the operation terminates successfully (line 65).
Otherwise, the next slab is loaded and we repeat the procedure.
}
% Deletion is very similar to our search procedure.
% Similarly, we use a boolean variable, \texttt{is\_active}, to indicate whether each thread has something to delete or not.
% We form a work queue, and one by one threads fetch the source key to be deleted at that round (similar to lines 2--7 in Fig.~\ref{alg:search}).
% Starting from the source key's bucket, we look for the key in each slab.
% If found, we mark that key and its value as deleted.
%%%%%%%%%%%%%%%%%%%%%%%%%%%%%%%%%%%%%%%%
\subsubsection{\textsc{Flush}}\label{subsubsec:flush_details}
Since we do not physically remove deleted elements in the slab hash but instead mark them as deleted, after a while it is possible to have slab lists that can be reorganized to occupy fewer slabs.
A \textsc{flush} operation takes a bucket as an argument and then a warp processes all slabs within that bucket's slab list and compacts them into fewer slabs.
In the end, we deallocate those emptied buckets in the SlabAlloc so that they can be reused by others.
In order to guarantee correctness, we implement this operation as a separate kernel call so that no other thread can perform an operation in those buckets while we are flushing its contents.
%%%%%%%%%%%%%%%%%%%%%%%%%%%%%%%%%%%%%%%%

%% file: code_example.tex
\begin{figure}
\ssmall
\algsetup{linenosize=\ssmall}
\begin{algorithmic}[1]
% \STATE {\textbf{\_\_device\_\_ void warp\_operation}(\begin{varwidth}[t]{\linewidth}
%         \hskip\algorithmicindent bool \&is\_active, \par
%         \hskip\algorithmicindent uint32\_t \&myKey, \par
%         \hskip\algorithmicindent uint32\_t \&myValue)\{
%         \end{varwidth}
% }
\revision{
\STATE {\textbf{\_\_device\_\_ void warp\_operation}(bool \&is\_active, uint32\_t \&myKey, uint32\_t \&myValue) \{ }

\STATE {next $\leftarrow$ BASE\_SLAB;}
\STATE {work\_queue $\leftarrow$ \_\_ballot(is\_active);}
\WHILE {(work\_queue != 0)}
\STATE {next $\leftarrow$ (if work\_queue is changed) ? (BASE\_SLAB) : next;}
\STATE {src\_lane $\leftarrow$ next\_prior(work\_queue); src\_key $\leftarrow$ \_\_shfl(myKey, src\_lane);}
% \STATE {src\_key $\leftarrow$ \_\_shfl(myKey, src\_lane);}
\STATE {src\_bucket $\leftarrow$ hash(src\_key); read\_data $\leftarrow$ ReadSlab(next, laneId);}
% \STATE {read\_data $\leftarrow$ ReadSlab(next, laneId);}
% \STATE {read\_data $\leftarrow$ (next == BASE\_ADDRESS)?(d\_bucket[src\_bucket * BASE\_UNIT\_SIZE + laneId]):(SlabAlloc(address\_decode(next) + laneId))}
\STATE {{warp\_search\_macro()} \textbf{OR} {warp\_replace\_macro()} \textbf{OR} {warp\_delete\_macro()}}
% \IF{{search operation}}
%       \STATE {warp\_search\_macro()}
% \ELSIF{{replace operation}}
%       \STATE {warp\_replace\_macro()}
% \ENDIF
\STATE {work\_queue $\leftarrow$ \_\_ballot(is\_active);}
\ENDWHILE

\STATE \}
}
\STATE {// ============================================}
\STATE {\textbf{warp\_search\_macro()}}
\STATE {found\_lane $\leftarrow$ \_\_ffs(\_\_ballot(read\_data == src\_key) \& VALID\_KEY\_MASK);}

\IF{(found\_lane is valid)}
        \STATE found\_value $\leftarrow$ \_\_shfl(read\_data, found\_lane + 1);
        \IF{(laneId == src\_lane)}
        \STATE {myValue $\leftarrow$ found\_value; is\_active $\leftarrow$ false;}
        % \STATE {is\_active $\leftarrow$ false;}
        \ENDIF
\ELSE
        \STATE next\_ptr $\leftarrow$ \_\_shfl(read\_data, ADDRESS\_LANE);
        \IF{(next\_ptr is an empty address pointer)}
                \IF{(laneId == src\_lane)}
                        \STATE {myValue $\leftarrow$ SEARCH\_NOT\_FOUND; is\_active $\leftarrow$ false;}
                        % \STATE is\_active $\leftarrow$ false;
                \ENDIF
        \ELSE
                \STATE next $\leftarrow$ next\_ptr;
        \ENDIF
\ENDIF
%%%%%%%%%%%%%%%%%%%%%%%%%%%%%%%%%%%%%%%%%%%%%%%%%%%%
\STATE {// ============================================}
\STATE {\textbf{warp\_replace\_macro()}}
\STATE {dest\_lane $\leftarrow$ {\tiny \_\_ffs(\_\_ballot(read\_data == EMPTY $\|$ read\_data == myKey) \& VALID\_KEY\_MASK)};}
\IF{dest\_lane is valid}
        \IF{(src\_lane == laneId)}
                \STATE {old\_pair $\leftarrow$ {\tiny atomicCAS(SlabAddress(next, dest\_lane), EMPTY\_PAIR, $\langle$ myKey, myValue $\rangle$)};}
                \IF{(old\_pair == EMPTY\_PAIR)}
                        \STATE{is\_active $\leftarrow$ false;}
                \ENDIF
        \ENDIF
\ELSE
        \STATE next\_ptr $\leftarrow$ \_\_shfl(read\_data, ADDRESS\_LANE);
        \IF{next\_ptr is empty}
                \STATE new\_slab\_ptr $\leftarrow$ SlabAlloc::warp\_allocate();
                \IF{(laneId == ADDRESS\_LANE)}
                        \STATE {temp $\leftarrow$ {\tiny atomicCAS(SlabAddress(next, ADDRESS\_LANE), EMPTY\_POINTER, new\_slab\_ptr)};}
                        \IF{(temp != EMPTY\_POINTER)}
                                \STATE SlabAlloc::deallocate(new\_slab\_ptr);
                        \ENDIF
                \ENDIF
        \ELSE
                \STATE {next $\leftarrow$ next\_ptr;}
        \ENDIF
\ENDIF
%%%%%%%%%%%%%%%%%%%%%%%%%%%%%%%%%%%%%%%%%%%%%%%%%%%%%%%%
\revision{
\STATE {// ============================================}
\STATE {\textbf{warp\_delete\_macro()}}
\STATE {dest\_lane $\leftarrow$ {\tiny \_\_ffs(\_\_ballot(read\_data == src\_key) \& VALID\_KEY\_MASK)};}
\IF{dest\_lane is valid}
        \IF{(src\_lane == laneId)}
                \STATE {*(SlabAddress(next, src\_lane)) $\leftarrow$ DELETED\_KEY;}
                \STATE {is\_active $\leftarrow$ false;}
        \ENDIF
\ELSE
        \STATE {next\_ptr $\leftarrow$ \_\_shfl(read\_data, ADDRESS\_LANE);}
        \IF{next\_ptr is empty}
                \STATE{is\_active $\leftarrow$ false;}
        \ELSE
                \STATE{next $\leftarrow$ next\_ptr;}
        \ENDIF
\ENDIF
}
%%%%%%%%%%%%%%%%%%%%%%%%%%%%%%%%%%%%%%%%%%%%%%%%%%%%%%%%
\end{algorithmic}
\caption{\revision{Pseudocode for search (\textsc{search}), insert (\textsc{replace}), and delete (\textsc{delete}) operations in the slab hash.}}
\vspace{-0.2in}
\label{alg:search}
\end{figure}

%% file: mem_alloc.tex
\section{Dynamic memory allocation}\label{sec:mem_alloc}
\paragraph{Motivation} Today's GPU memory allocators
(Section~\ref{sec:related}) are generally designed for variable-sized allocations and aim to avoid too much memory fragmentation (so as not to run out of memory with large allocations). These allocators are designed for flexibility and generality at the cost of high performance; for instance, they do not emphasize branch and memory-access divergence. For high-throughput mutability scenarios such as hash table insertions (e.g., the slab hash) that require many allocations, the memory allocator would be a significant bottleneck.

The WCWS strategy (Section~\ref{subsec:warp_focus}) that we chose for the slab hash results in the following allocation problem: \revision{insertion operations that are assigned to a single warp are sequentially processed (one at a time) and hence we will require numerous independent fixed-size slab allocations per warp at different times during a warp's lifetime.
These allocations cannot be simply formed into a single larger coalesced allocation that suits other allocators.}

Consequently, current allocators perform poorly on this pattern of allocations. For example, on a Tesla K40c (ECC disabled), with one million slab allocations, 128 bytes per slab, one allocation per thread and with similar total used memory for each allocator, CUDA's \texttt{malloc} spends 1.2s (0.8~M slabs/s). \texttt{Halloc} takes 66 ms (16.1~M slabs/s).
We designed our own memory allocator that is better suited for this allocation workload.
Our \texttt{SlabAlloc} takes 1.8~ms (600~M slabs/s), which is about 37x faster than \texttt{Halloc}.
% \Saman{Will add other allocator's soon, it needs more work to be able to use their code.} \John{Not high priority.}

\paragraph{Terminology} We use a hierarchical memory structure as follows: several ($N_S$) memory pools called \emph{super blocks} are each divided into several ($N_M$) smaller \emph{memory blocks}. Each memory block consists of fixed $N_U = 1024$ \emph{memory units} (i.e., slabs).
\remove{Figure~\ref{fig:alloc_mem_layout} shows this hierarchy.

\begin{figure}
\centering
\includegraphics[width = 1\linewidth]{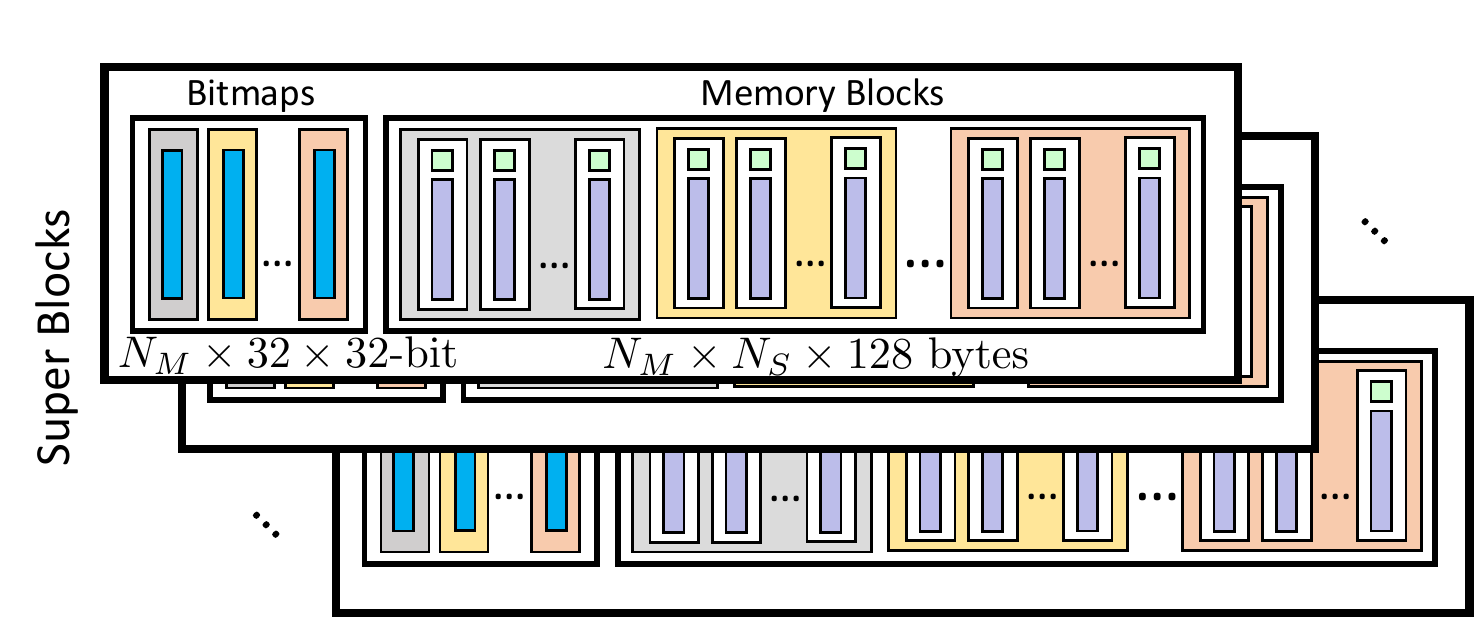}
\caption{Memory layout for SlabAlloc.}
\vspace{-0.2in}
\label{fig:alloc_mem_layout}
\end{figure}
}
\paragraph{SlabAlloc}
There are a total of $N_SN_M$ memory blocks. We distribute memory blocks uniformly among all warps, such that different warps may be assigned to each memory block. We call each warp's assigned memory block \emph{resident}.
A resident block is used for all allocations requested from its warp owners for up to 1024 memory units (slabs).
Once a resident block gets full, its warp randomly chooses (with a hash) another memory block and uses that as its new resident block.
After a threshold number of resident changes, we add new super blocks and reflect them in the hash functions.
In its most general case, both the super block and its memory block are chosen randomly using two different hash functions (taking the global warp ID and the total number of resident change attempts as input arguments).
This creates a probing effect in the way we assign resident blocks.
Since there are 1024 memory units within each memory block, by using just one 32-bit bitmap variable per thread (a total of 32$\times$32-bit across the warp), a warp can fully store a memory block's full/empty availability.

Upon each allocation request, all threads in the warp look into their local resident bitmap (stored in a register) and announce whether there are any unused memory units in their portion of the memory block. For example, thread 0 is in charge of the first 32 memory units of its resident block, thread 1 has memory units 32--63, etc.
Following a pre-defined priority order (e.g., the least indexed unused memory unit), all threads then know which thread should allocate the next memory unit.
That thread uses an atomicCAS operation to update its resident bitmap in global memory.
If successful, then the newly allocated memory unit's address is shared with all threads within that warp (using shuffle instructions).
If not, it means some other warp has previously allocated new memory units from this memory block and the local register-level resident bitmap should be updated.
As a result, in the best case scenario and with low contention, each allocation can be addressed with just a single atomic operation.
If necessary, each resident change requires a single coalesced memory access to read all the bitmaps for the new resident block.
Deallocation is done by first locating the slab's memory block's bitmap in global memory and then atomically unsetting the corresponding bit.

% \John{Maybe add another \texttt{paragraph} that notes the typical cost of an allocation, i.e., what usually happens, e.g., ``A typical allocation request finds a free memory unit within the current resident memory block. This costs \ldots'' (e.g., one atomic, one global memory read, etc.). You can contrast here against previous work. You can also put this at the end of the section because you'll have a different cost for the two allocators.}\Saman{I guess I have talked about some of these points in previous paragraph. Will add more in-depth description in the future.}

% \John{Might be useful to have a \texttt{paragraph} on \texttt{Memory layout}. This can cover two things: what you ask CUDA's \texttt{malloc} to do and what the restrictions are on the layout (e.g., that super blocks are allocated as a single contiguous array).}\Saman{Talked about this in the beginning of SlabAlloc-light introduction.}

\paragraph{Memory structure}
In general, we need a 64-bit pointer variable to uniquely address any part of the GPU's memory space.
Almost all general-purpose memory allocators use the same format.
Since our main target is to improve our data structure's dynamic performance, we trade off the generality of our allocators to gain performance: we use 32-bit \emph{address layouts}, which are less expensive to store and share (especially because shuffle instructions work only with 32-bit registers).
In order to uniquely address a memory unit, we use a 32-bit variable: 1) the first 10 bits represent the memory unit's index within its memory block, 2) the next 14 bits are used for the memory block's index within its super block, and 3) the next 8 bits represent the super block.
Each super block is assumed to be allocated continuously on a single array ($<$ 4~GB).
\remove{Each memory unit is at least $2^7$ bytes, and there are total of 1024 units in each memory block.
Considering $2^7$ bytes per each block's bitmap, we can at most put $2^{14}$ memory blocks within each super block (i.e., $(2^7 + 2^{17})N_M \leq 2^{32} \Rightarrow N_M < 2^{15}$).
As a result, with this version we can dynamically allocate memory up to $2^7N_SN_MN_U < $~1~TB (much larger than any current GPU's DRAM size).}
% \John{If you have room, a figure to show the layout for both 32b and 64b allocators would be helpful.}\Saman{I don't have any 64-bit allocator implemented, at least for the latest optimized versions that I used. It requires twice as many number of shuffle instructions, that I didn't like from the first place. But, will spend more time on this in the future to have some meaningful comparison.}

\remove{In SlabAlloc, we assume that each super block is allocated as a contiguous array.
In order to look up allocated slabs using our 32-bit layout, we store the actual beginning address (64-bit pointers) of each super block in shared memory.
Before each memory access, we must first decode the 32-bit layout variables into an actual 64-bit memory address.
This requires a single shared memory access per memory lookup (using the above 32-bit variable), which is costly, especially when performing search queries.
To address this cost, we can also implement a lightweight memory allocator, \emph{SlabAlloc-light}, where all super blocks are allocated in a single contiguous array. In this case, a single beginning address for the first super block, which is stored as a global variable, is enough to find the address of all memory units, resulting in a less expensive memory lookup, but with less scalability (at most about 4~GB).
% \John{Also perhaps useful is an example, say for the $2^{22}$-element hash table in Figure~\ref{fig:bulk_lf}: this many super blocks, this many memory blocks, 1024 memory units, and (maybe?) an overall size overhead due to the bitmaps of $N$\%. Maybe pick two examples, one that uses the 64b allocator and one that uses the 32b allocator.}
In scenarios where memory lookups are heavily required (e.g., the bulk search scenarios in Section~\ref{subsec:bulk}), SlabAlloc-light gives us up to 25\% performance improvement compared to the regular SlabAlloc.
}

%% file: perf_eval.tex
\section{Performance Evaluation}\label{sec:perf_eval}
We evaluate our slab list and slab hash on an NVIDIA Tesla K40c GPU (with ECC disabled), which has a Kepler microarchitecture with compute capability of 3.5, 12~GB of GDDR5 memory, and a peak memory bandwidth of 288~GB/s.
We compile our codes with the CUDA 8.0 compiler (V8.0.61).
In this section, all our insertion operations maintain uniqueness (\textsc{replace}).
We have also used \replace{SlabAlloc-light with 32 super blocks}{SlabAlloc with 32 super blocks (on a contiguous allocation)}, 256 memory blocks, and 1024 memory units, 128 bytes each.
We believe this is a fair comparison, because all other methods that we compare against (CUDPP and Misra's) pre-allocate a single contiguous array for use. If necessary, our SlabAlloc can be scaled up to 1~TB allocations.
%%%%%%%%%%%%%%%%%%%%%%%%%%%%
% \subsection{Slab Hash}\label{subsec:perf_slab_hash}
% To the best of the authors' knowledge, there are no other available dynamic hash tables for GPUs.
We divide our performance evaluations into two categories. First, we compare against other static hash tables (such as CUDPP's cuckoo hashing implementation~\cite{Alcantara:2009:RPH:nourl}) in performing operations such as building the data structure from scratch and processing search queries afterwards.
\revision{Alcantara did an extensive study over various GPU hash tables including linear and quadratic probing methods and reached the conclusion that their cuckoo hashing implementation was substantially superior~\cite{Alcantara:2011:EHT}.}
Second, we design a concurrent benchmark to evaluate the dynamic behavior of our proposed methods with a random mixture of certain operations performed asynchronously (insertion, deletion, and search queries). We compare the slab hash to Misra and Chaudhuri's lock-free hash table~\cite{Misra:2012:PEC}.
%%%%
\subsection{Bulk benchmarks (static methods)}\label{subsec:bulk}
There are two major operations defined for static hash tables such as CUDPP's hash table:
(1) building the data structure given a fixed load factor (i.e., memory utilization) and an input array of key-value pairs, and (2) searching for an array of queries (keys) and returning an array of corresponding values (if found).
By giving the same set of inputs into our slab hash, where each thread reads a key-value pair and dynamically inserts it into the data structure, we can build a hash table.
Similarly, after the hash table is built, each thread can read a query from an input array and search for it dynamically in the slab hash and store back the search results into an output array. By doing so, we can compare slab hash with other static methods.%
\remove{\footnote{In the slab hash, there is no difference between a bulk build operation and incremental insertions of a batch of key-value pairs. However, for a bulk search we assign more queries to each thread.}}

For many data structures, the performance cost of supporting incremental mutability is significant: static data structures often sustain considerably better bulk-build and query rates when compared to similar data structures that additionally support incremental mutable operations. We will see, however, that the performance cost of supporting these additional operations in the slab hash is modest.

\begin{figure}
\centering
\subfloat[Building rate]{
        \includegraphics[width=0.49\linewidth]{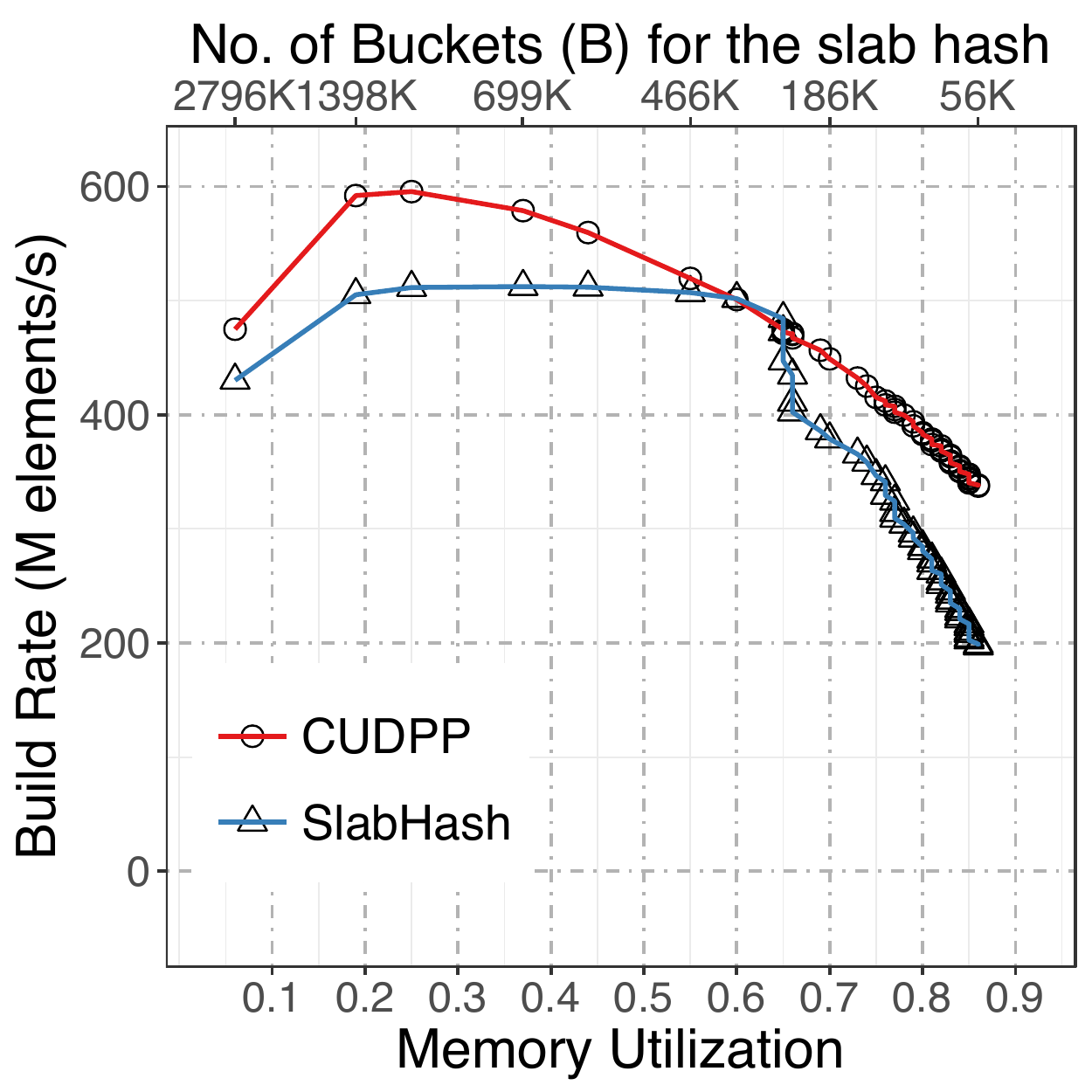}
        \label{fig:bulk_build_lf}
}
\subfloat[Search rate]{
        \includegraphics[width=0.49\linewidth]{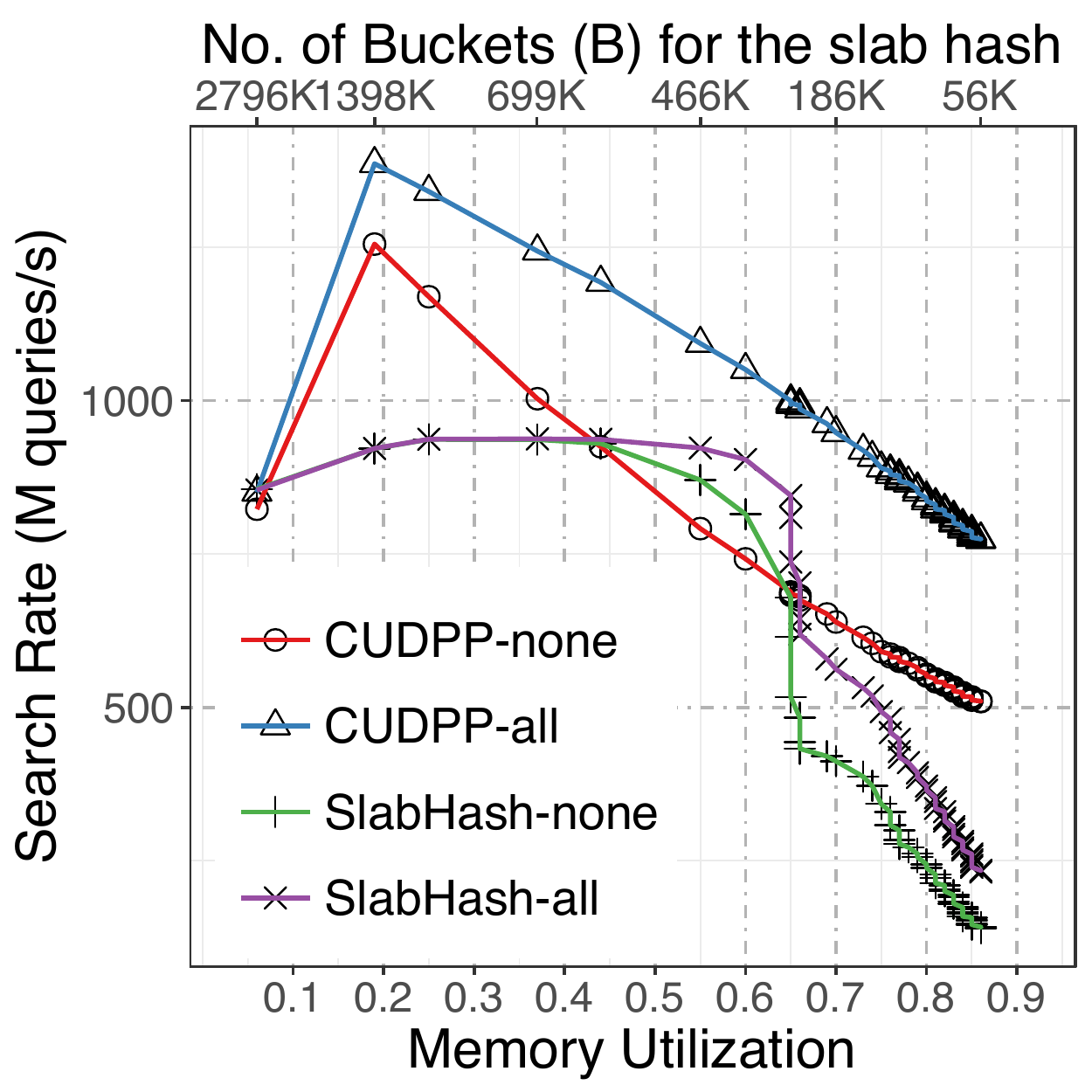}
        \label{fig:bulk_search_lf}
}\\
\subfloat[\revision{Memory utilization versus average slab count ($\beta$)}]{
        \includegraphics[width=0.8\linewidth]{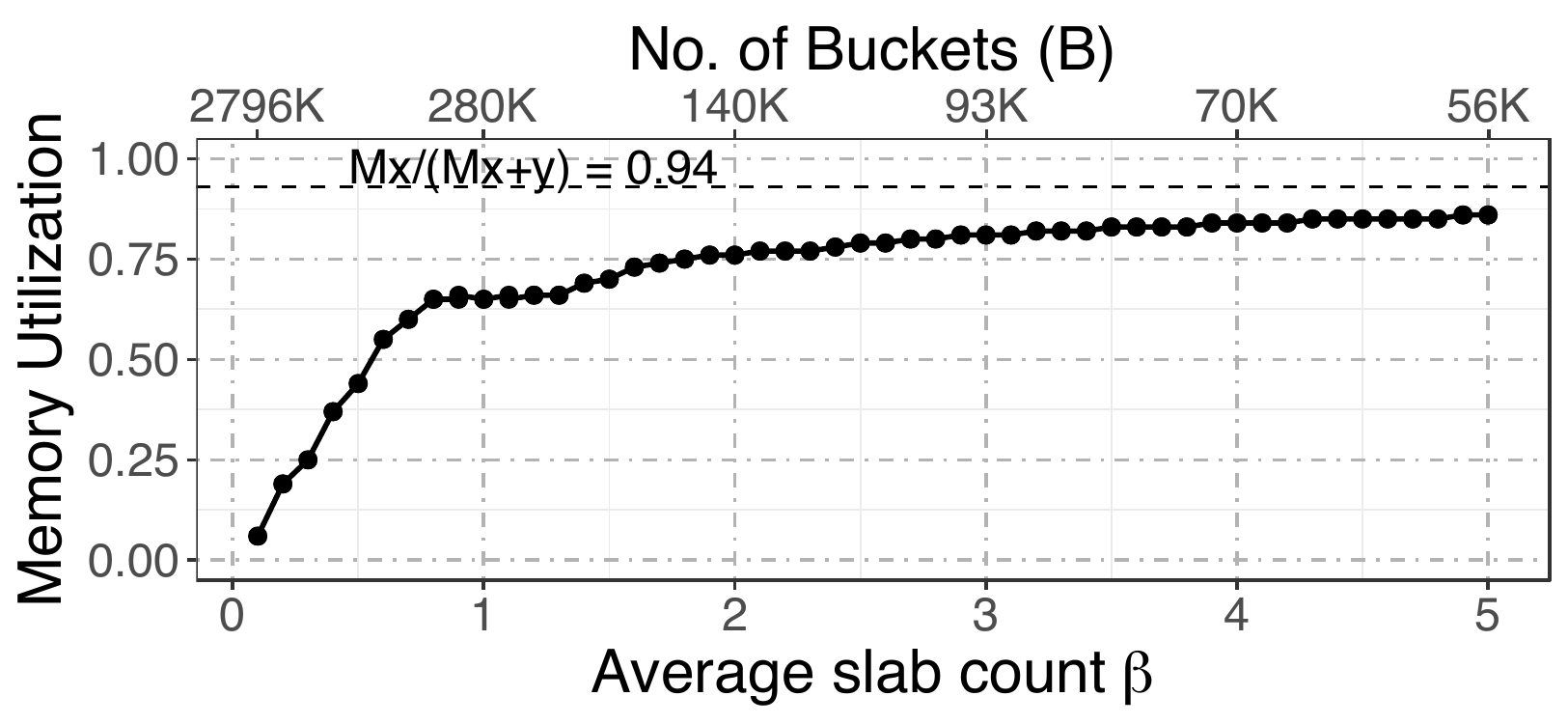}
        \label{fig:bucket_plot}
}
\caption{\small \revision{Performance (M operations/s) versus memory efficiency.
$2^{22}$ elements are stored in the hash table in each trial.
On top, the number of buckets for the slab hash is shown.
(a) The whole table is built from scratch, dynamically, and in parallel.
(b) There are $2^{22}$ search queries where all (or none) of them exist.
(c) Achieved memory utilization versus the average slab count (and number of buckets) is shown.}
}
\label{fig:bulk_lf}
\end{figure}

Figure~\ref{fig:bulk_build_lf} shows the build rate (M elements/s) for various memory utilizations.
$n=2^{22}$ elements are stored in the table.
For CUDPP's hash table, memory utilization (load factor) can be directly fixed as an input argument, but the situation is slightly more involved for the slab hash:
\revision{Given a fixed number of buckets ($B$), the average slab count is $\beta = n/(MB)$, where $M$ is the number of elements per slab as defined in Section~\ref{subsec:hash_table}.
The performance of the slab hash and its achieved memory utilization is directly affected by the number of buckets (and $\beta$). Figure~\ref{fig:bucket_plot} shows the achieved memory utilization vs.\ the average slab count and number of buckets.
So, on average, in order to achieve a particular memory utilization we can refer to Fig.~\ref{fig:bucket_plot} and choose the optimal $\beta$ and then compute the required number of initial buckets.
The maximum memory utilization (given our choice of parameters for the slab hash) is about 94\%, which is achieved as $B \to 1$.
}
% \revision{By choosing various average slab counts (from 0.1 to 5.0 in Fig.~\ref{fig:bulk_build_lf}), we can compute the initial number of buckets (from 2.8~M to 55~k buckets respectively).
% As a result, for each data point in the figure, we first build the slab hash and compute its memory utilization (the total size of stored elements divided by the total used memory), and then build a CUDPP hash with the same utilization and with the same input elements.}

In our simulations, then,  we build a CUDPP hash with the same utilization and with the same input elements.
The process is averaged over 50 independent randomly generated trials.
For search queries on a hash table with $n=2^{22}$ elements, we generate two sets of $n = 2^{22}$ random queries: 1) all queries exist in the data structure; 2) none of the queries exist. These two scenarios are important as they represent, on average, the best and worst case scenarios respectively.
Figure~\ref{fig:bulk_search_lf} shows the search rate (M queries/s) for both scenarios with various memory utilizations.

The slab hash gets its best performance from 19--60\% memory utilization; these utilizations have a 0.2--0.7 average slab count.
Intuitively, this is when the average list size fits in a single slab.
% \footnote{To choose the optimal number of initial buckets, we can choose the preferred memory utilization and its corresponding average slab count $\beta$ from Fig.~\ref{fig:bulk_lf}. Then $B = n/(M\beta)$, where $M = 15$, as discussed in Section~\ref{subsec:hash_table} and \ref{subsec:parameters}.}
Peak performance is 512~M insertion/s and 937~M queries/s.
At about 65\% memory utilization there is a sudden drop in performance for both insertions and search queries.
This drop happens when the average slab count is around 0.9--1.1, which means that almost all buckets will have more than one slab and most of the operations will have to visit the second slab.
The slab hash appears to be competitive to cuckoo hashing, especially around 45--65\% utilization.
For example, our slab hash is marginally better in insertions (at 65\%) and 1.1x faster in search queries when no queries exist.
But, using a geometrical mean over all utilizations and $n=2^{22}$, cuckoo hashing is 1.33x, 2.08x, and 2.04x faster than the slab hash for build, search-all, and search-none respectively.
% \Saman{todo: CUDPP's peak.}

\begin{figure}
\centering
\subfloat[Building rate]{
        \includegraphics[width=0.49\linewidth]{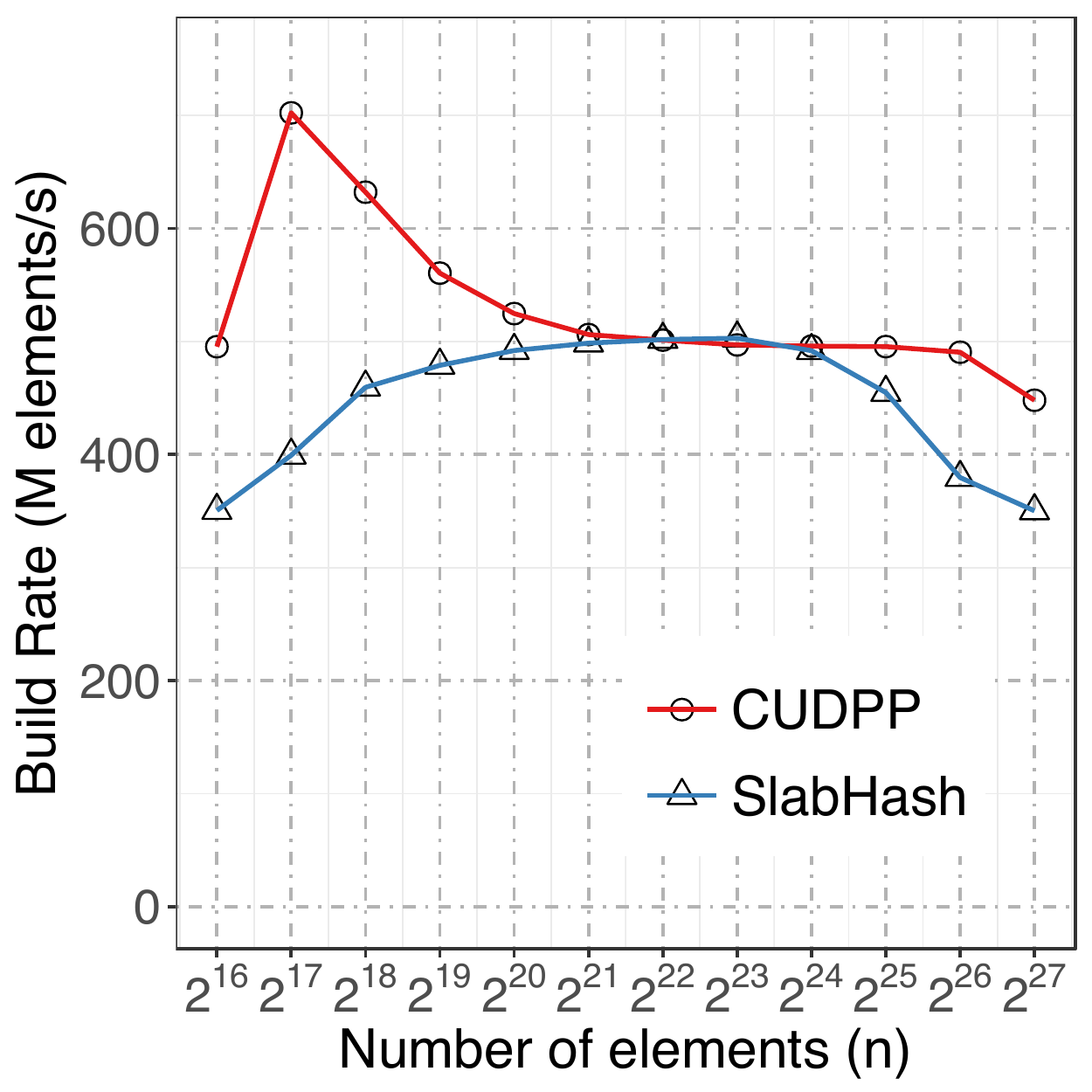}
}
\subfloat[Search rate]{
        \includegraphics[width=0.49\linewidth]{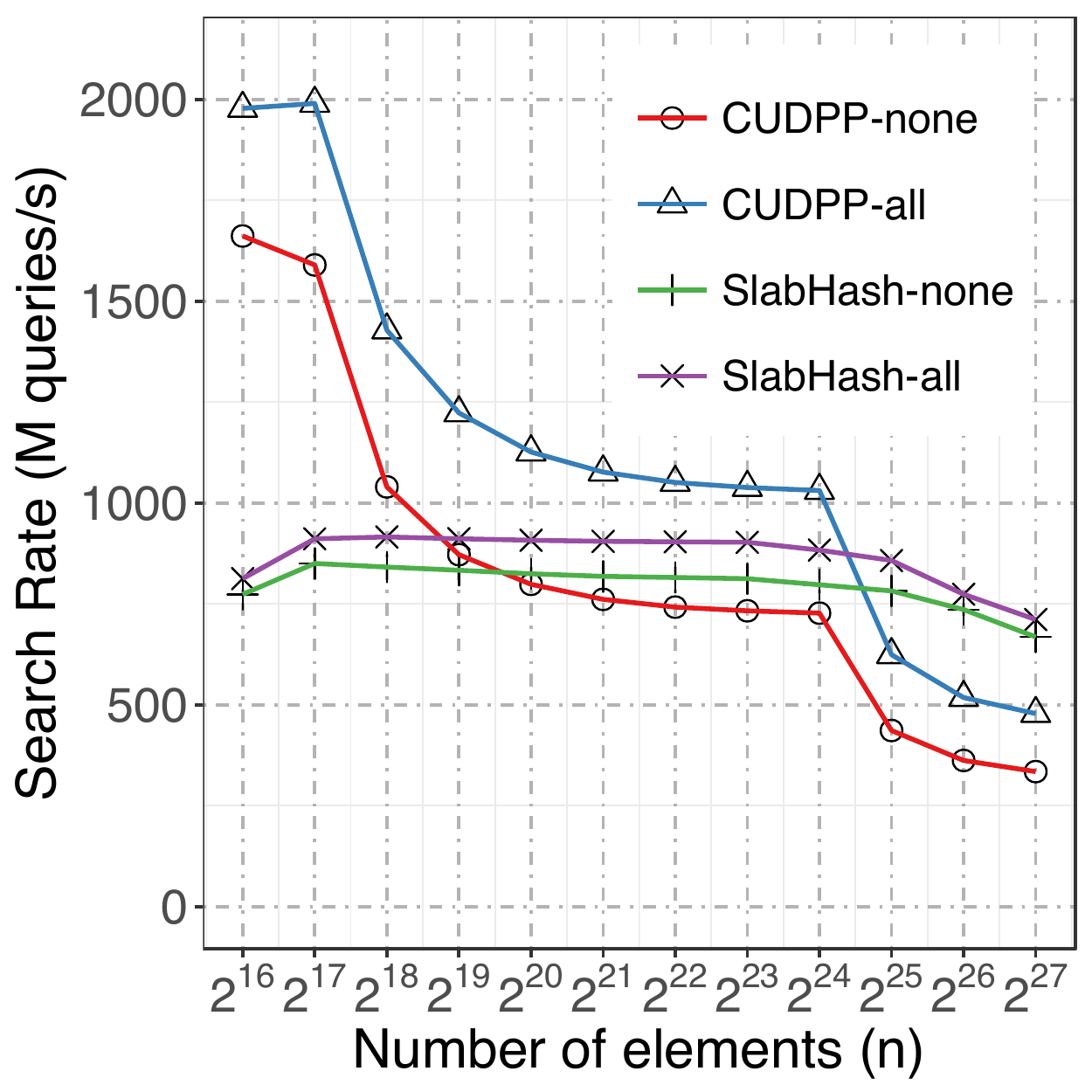}
}
\label{fig:bulk_table_size}
\caption{\small Performance (M operations/s) versus total number of stored elements in the hash table.
Memory utilization is fixed to be 60\%.
(a) The whole table is built from scratch, dynamically, and in parallel.
(b) There are as many search queries as there are elements in the table where either all (or none) of them exist.
}
\label{fig:bulk_static}
\end{figure}

Figure~\ref{fig:bulk_static} shows the build rate (M elements/s) and search rate (M queries/s) vs.\ the total number of elements ($n$) stored in the hash table, where memory utilization is fixed to be 60\% (an average slab count of 0.7).
Here we witness that CUDPP's building performance is particularly high when the table size is small, which is because most of the atomic operations can be done in cache level.
The slab hash saturates the GPU's resources for $2^{20} \leq n \leq 2^{24}$, where both methods perform roughly the same.
For very large table sizes, both methods degrade, but the slab hash's performance decline starts at smaller table sizes.
For search queries, the slab hash shows a relatively consistent performance with a harmonic mean of 861 and 793 M~queries/s for search-all and search-none.
In this experiment, with a geometric mean over all table sizes and 65\% memory utilization, the speedup of CUDPP's cuckoo hashing over the slab hash is 1.19x, 1.19x, and 0.94x for bulk build, search-all, and search-none respectively.
% \Saman{CUDPP's peak and fall for search} \John{It's confusing to see this result and also the similar but different-numbers one at the end of the previous paragraph.}

Ideally, the ``fast path'' scenario for CUDPP's cuckoo hash table requires a single atomicCAS for insertion and a single random memory access for a search.
Unless there is some locality to be extracted from input elements (which does not exist in most scenarios), any hash table is doomed to have at least one global memory access (atomic or regular) per operation.
This explains why CUDPP's peak performance is hard to beat, and other proposed methods such as stadium hashing~\cite{Khorasani:2015:SHS} and Robin Hood hashing~\cite{Garcia:2011:CPH} are unable to compete with its peak performance.
In the slab hash, for insertion, ideally we will have one memory access (reading the slab) and a single atomicCAS to insert into an empty lane.
For search, it will be a single memory access plus some overhead from extra warp-wide instructions (Section~\ref{sec:impl_details}).
%% Experiment for bulk versus incremental:
\begin{figure}
\centering
\includegraphics[width=0.73\linewidth]{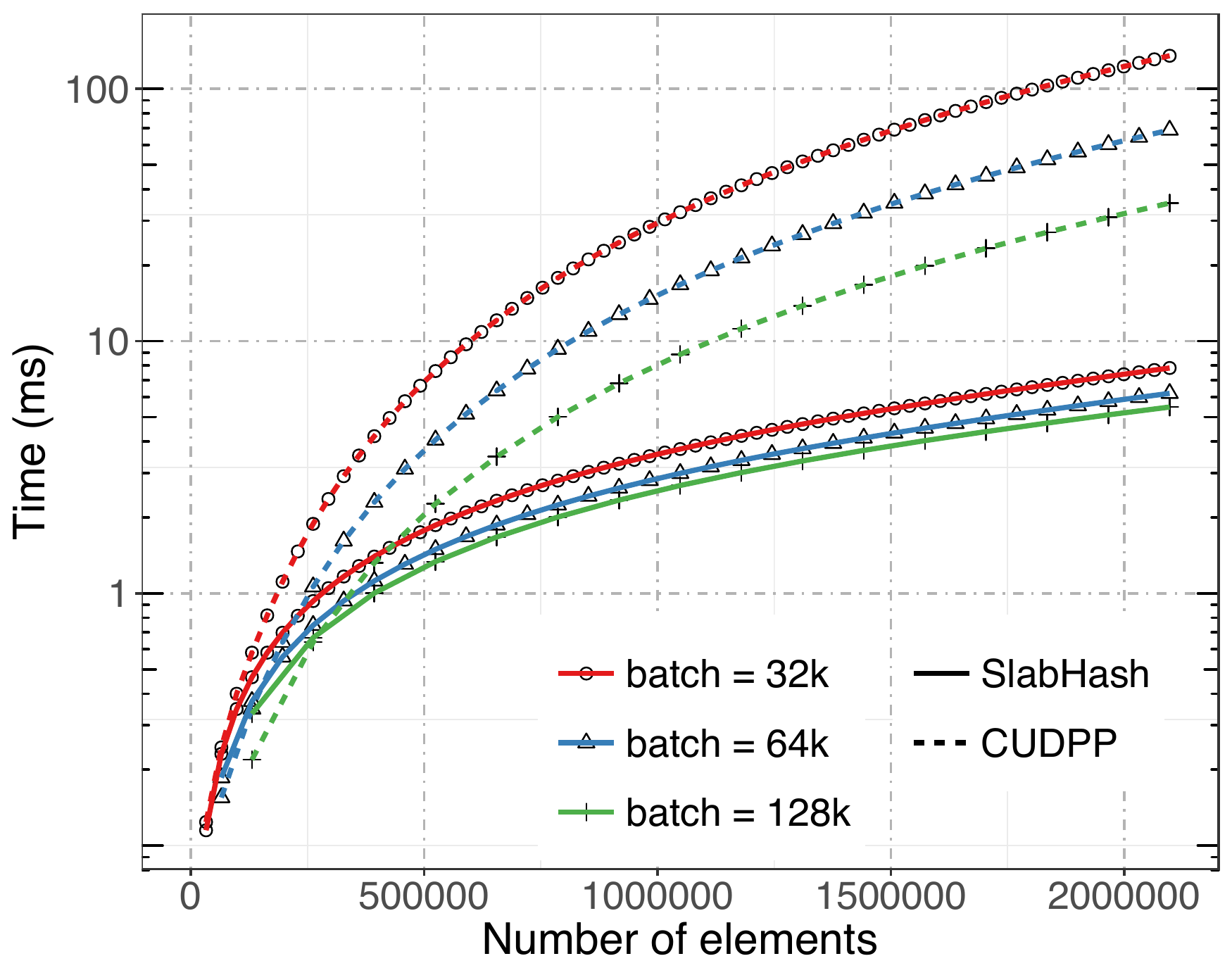}
\caption{\small Incremental batch update for the slab hash, as well as building from scratch for the CUDPP's cuckoo hashing. Final memory utilization for both methods are fixed to be 65\%. Time is reported in logarithmic scale.}
\label{fig:incremental_batch}
\end{figure}

\subsection{Incremental insertion}
Suppose we periodically add a new batch of elements to a hash table. For CUDPP, this means building from scratch every time.
For the slab hash, this means dynamically inserting new elements into the same data structure.
Figure~\ref{fig:incremental_batch} shows both methods in inserting new batches of different sizes (32k, 64k, and 128k) until there are 2 million elements stored in the hash table.
For CUDPP, we use a fixed 65\% load factor.
For the slab hash, we choose initial number of buckets so that its final memory utilization (after inserting all batches) is 65\%.
As expected, the slab hash significantly outperforms cuckoo hashing by reaching final speedup of 6.4x, 10.4x, and 17.3x for batches of size 128k, 64k, and 32k.
As the number of inserted batches increases (as with smaller batches), the performance gap increases.

\subsection{Concurrent benchmarks (dynamic methods)}\label{subsec:concurrent}
One notable feature of the slab hash is its ability to perform truly concurrent query and mutation (insertion/deletion) operations without having to divide different operations into different computation phases.
To evaluate our concurrent features, we design the following benchmark.
% To evaluate the effectiveness of our dynamic hash table in handling different operations concurrently, we design the following benchmark.
Suppose we build our hash table with an initial number of elements.
We then continue to perform operations in one of the following four categories:
a) inserting a new element, b) deleting a previously inserted element, c) searching for an existing element, d) searching for a non-existing element.
We define an \emph{operation distribution} $\Gamma = (a,b,c,d)$, such that every item is non-negative and $a+b+c+d = 1$.
Given any $\Gamma$, we can construct a random workload where, for instance, $a$ denotes the fraction of new insertions compared to all other operations.
To ensure correctness, we generate operations in batches and process batches one at a time, but each in parallel.
For each batch, operations are randomly assigned to each thread (one operation per thread) such that all four operations may occur within a single warp.
In the end, we average the results over multiple batches.
We consider three scenarios: 1)~$\Gamma_0=(0.5, 0.5, 0, 0)$ where all operations are updates, 2)~$\Gamma_1 = (0.2, 0.2, 0.3, 0.3)$ where there are 40\% updates and 60\% search queries, and 3)~$\Gamma_2 = (0.1, 0.1, 0.4, 0.4)$ where there are 20\% updates and 80\% search queries.

\begin{figure}
  \vspace{-0.2in}
\centering
\subfloat[Concurrent benchmark]{
                \includegraphics[width = 0.47\linewidth]{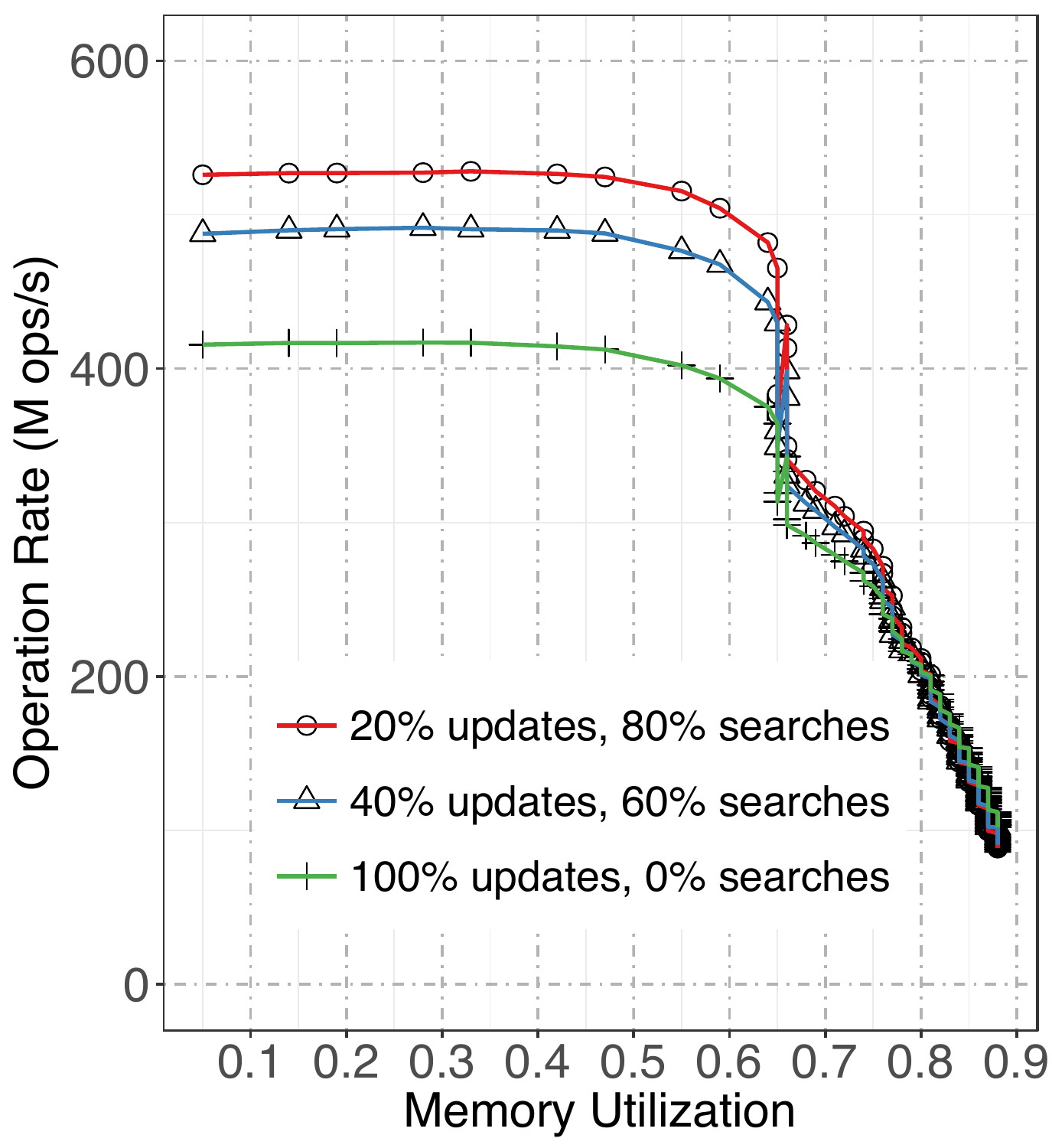}%
                \label{fig:concurrent_slab}
}
\subfloat[Performance vs.\ Misra's]{
                \includegraphics[width = 0.49\linewidth]{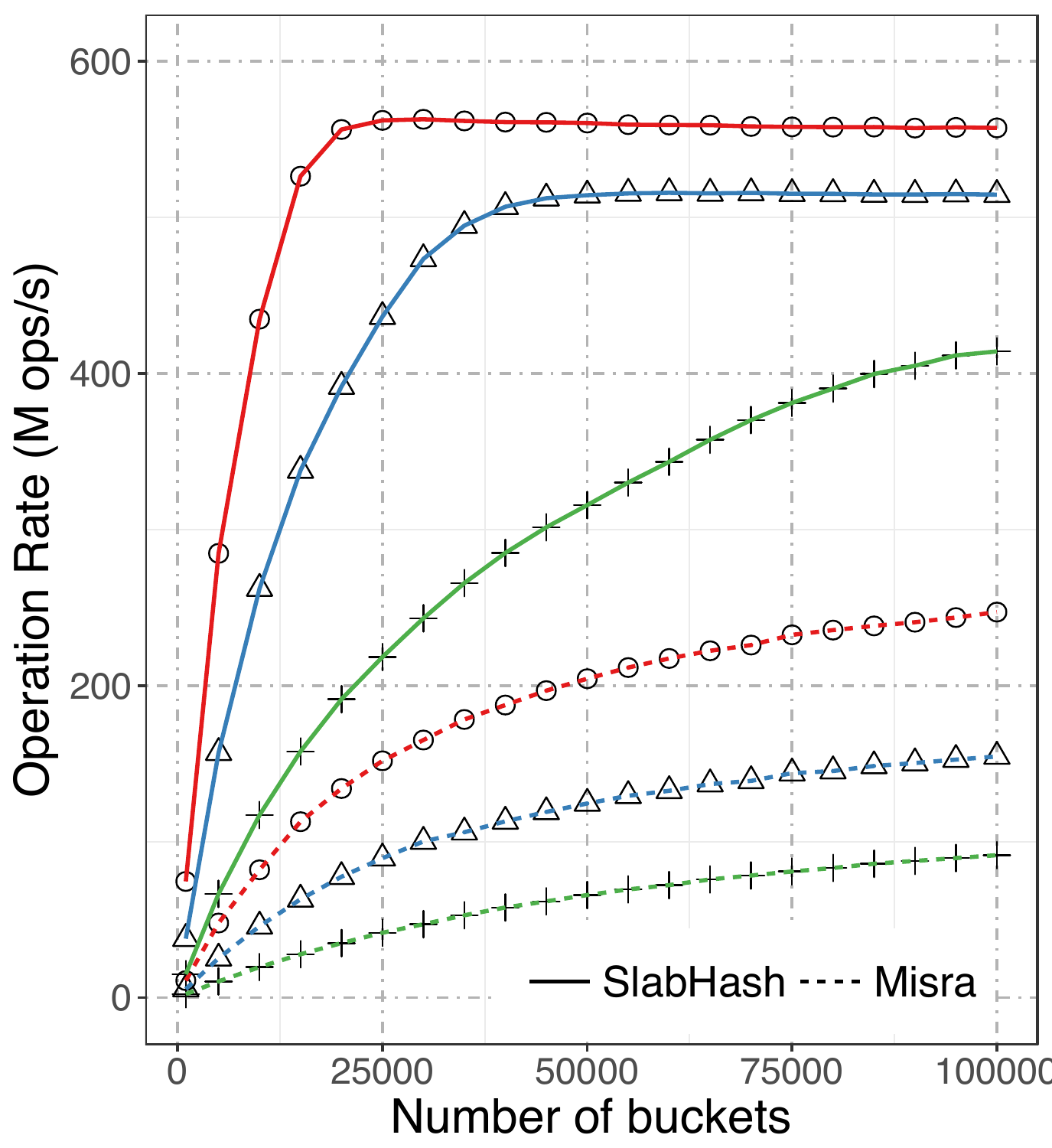}%
                \label{fig:concurrent_misra}
}
\caption{\small (a) Concurrent benchmark for the slab hash: performance (M ops/s) versus initial memory utilization. (b) Performance (M ops/s) versus Misra and Chaudhuri's lock-free hash table~\cite{Misra:2012:PEC}.
Three different operation distributions are shown in different colors as shown in (a).\vspace{-0.2in}}
\label{fig:concurrent_bench}
\end{figure}

Figure~\ref{fig:concurrent_slab} shows the slab hash performance (M~ops/s) for three different operation distributions and various initial memory utilizations.
Since updates are computationally more expensive than searches, given a fixed memory utilization, performance gets better with fewer updates ($\Gamma_0 < \Gamma_1 < \Gamma_2$).
Similar to the behavior in Fig.~\ref{fig:bulk_lf}, the slab hash sharply degrades in performance with more than 65\% memory utilization, falling to about 100~M ops/s with about 90\% utilization.
Comparing against our bulk benchmark in Fig.~\ref{fig:bulk_lf}, it is clear that the slab hash performs slightly worse in our concurrent benchmark (e.g., $\Gamma_0$ in Fig~\ref{fig:concurrent_slab} and Fig.~\ref{fig:bulk_build_lf}).
There are two main reasons: (1) Since it is assumed that in static situations all operations are available, we can assign multiple operations per thread and hide potential memory-related latencies, and (2) in concurrent benchmarks we run three different procedures (one for each operation type) compared to the bulk benchmark that runs just one.

\paragraph{Misra's hash table} Misra and Chaudhuri have implemented a lock-free hash table using classic linked lists~\cite{Misra:2012:PEC}.
This is a key-only hash table (i.e., an unordered set), without any pointer dereferencing or dynamic memory allocation; based on the required number of insertions, an array of linked list nodes are allocated at compile time, and then indices of that array are used in the linked lists.
Since it uses a simplified version of a classic linked list (32-bit keys and 32-bit \emph{next} indices), it theoretically can reach at most 50\% memory utilization.
In order to compare its performance with our slab hash, we use our concurrent benchmarks and the three operation distributions discussed above. Figure~\ref{fig:concurrent_misra} shows performance (M ops/s) versus number of buckets, where each case has exactly one million operations to perform.
The slab hash significantly outperforms Misra's hash table, with geometric mean speedup of 5.1x, 4.3x, and 3.1x for distributions with 100\%, 40\% and 20\% updates respectively.

\remove{As discussed in Section~\ref{sec:related}, Moscovici et al.\ has recently proposed a lock-based skip list (GFSL). On a GeForce GTX 970, with 224 GB/s memory bandwidth, they report that its peak performance is about 100~M queries/s for searches and 50~M updates/s for updates (compared to our peak results of 937 and 512~M op/s respectively).
In the best case, GFSL requires at least two atomic operations (lock/unlock) and two other regular memory accesses for a single insertion.
This cost makes it unlikely that GFSL can outperform static cuckoo hashing (1 atomic/insert) or our dynamic slab hash (1 read and 1 atomic per insert) in their peak performance.}

%% file: conclusion.tex
\section{Conclusion}\label{sec:conclusions}
The careful consideration of GPU hardware characteristics as well as our warp-cooperative work sharing strategy lead us to design and implementation of an efficient dynamic hash table for GPUs.
Beyond getting significant speedup compared to previous semi-dynamic hash tables, slab hash proves to be competitive to the fastest static hash tables too.
We believe our slab list design and its utilization in slab hash can be a promising first step to provide a larger family of dynamic data structures with specialized analytics for them, 
which can also be used to target other interesting problems such as sparse data representation and dynamic graph analytics.    

%% file: ack.tex
\section*{Acknowledgments}\label{sec:ack}
Thanks to NVIDIA who provided the GPUs that made this research possible.
We appreciate the funding support from a 2016--17 NVIDIA Graduate Fellowship, from NSF awards CCF-1637442, CCF-1629657, OAC-1740333, CCF-1724745, CCF-1715777, CCF-1637458, and IIS-1541613, from the Defense Advanced Research Projects Agency (DARPA), from an Adobe Data Science Research Award, and from gifts from EMC and NetApp.